\documentstyle[amssymb,12pt]{article}

\begin{document}

\title{ON  NON-COMMUTATIVE INTEGRABLE BURGERS EQUATIONS}
\author{ Metin G{\" u}rses\\
{\small Department of Mathematics, Faculty of Sciences}\\
{\small Bilkent University, 06800 Ankara - Turkey}\\
\\
Atalay Karasu   \\
{\small Department of Physics, Faculty of Arts and  Sciences}\\
{\small Middle East Technical University, 06531 Ankara-Turkey}\\
\\
Refik Turhan\\
{\small Department of Engineering Physics}\\
{\small Ankara University, 06500 Ankara-Turkey}}
\begin{titlepage}
\maketitle

\begin{abstract}
We construct the recursion operators for the non-commutative
Burgers equations using their Lax operators. We investigate the
existence of any integrable mixed version of left- and
right-handed Burgers equations on higher symmetry grounds.
\vspace{0.5 in}
\end{abstract}

\end{titlepage}
\section{Introduction}

Non-commutative generalizations of the classical nonlinear
evolution equations in (1+1) dimensions were classified according
to the  symmetry based integrability in \cite{sak}. In this
classification, (besides some multi-component equations)
one-component non-commutative versions of a Korteweg-de Vries
(ncKdV), a Potential KdV (ncPKdV), two Modified KdV (ncMKdV1,
ncMKdV2) and two (left- and right-handed) nc-Burgers equations are
observed to have  higher symmetry in a certain weighting scheme of
symmetries. Recursion operators for ncKdV, ncPKdV and ncMKdV1 were
given in \cite{olv} and the one for ncMKdV2 in \cite{m1}. In
\cite{olv}, classification \cite{sak} is shown to be complete
under the assumed weighting scheme of equations.

Possession of a higher symmetry is a necessary condition which needs to be supplemented
by either a recursion operator or a  master symmetry for an equation to be integrable
in the symmetry sense. A recursion operator generates infinite hierarchy of symmetries
by mapping a symmetry to another endlessly. Master symmetries do the same as adjoint action.
Therefore ncKdV, ncPKdV, ncMKdV1 and ncMKdV2 are integrable equations.

As for the nc-Burgers equations, their integrability is proven by
a master symmetry \cite{olv}. Moreover,  they  are shown  to be
linearizable by a non-commutative version of the Cole-Hopf
transformation. Their symmetry hierarchies are obtainable from the
`higher heat equations' and they admit auto-B{\"a}cklund
transformations \cite{kup1,kup2}. Some exact solutions of the
associated linearized (nc-Heat) equation were obtained in
\cite{mar,ham}. So, nc-Burgers equations are one of the best
studied equations among the known non-commutative integrable
equations. There is one missing point however; currently no
recursion operator of nc-Burgers equations is known. It is even
claimed that there exists none \cite{olv}.

Integrability has many aspects. A symmetry integrable equation may further
possess a hierarchy of conservation laws, or a Lax formulation, auto-B{\"a}cklund transformations,
Hirota bilinear formulation, Painlave property etc. In general,
there is not a well established correspondence among these structures.
But demonstration of one of these structures is regarded as a strong indication
(and a good motivation to search) for other structures of an equation.

Recently, a new  nc-Burgers equation, which is a particular
parametric mixture of the left- and right-handed nc-Burgers
equations, was introduced with a Lax pair in \cite{ham}. Despite
having a Lax formulation, absence of this mixed nc-Burgers
equation in the former symmetry and the relevant structure studies
is remarkable.

In this paper, starting from a Kupershmidt type hierarchy of Lax
representations, we construct both the time independent and the
time dependent recursion operators of the  nc-Burgers equations by
the method introduced in \cite{m1}. Our construction leads to only
recursion operators of the left- or right-handed nc-Burgers
equations . The Lax representation given for the mixed version of
nc-Burgers equation does not lead to a recursion operator of the
equation  by the mentioned technique. Therefore, we reinvestigate
the possibility of having an integrable mixed version of
nc-Burgers equation on higher symmetry grounds again. This time,
however, we do the symmetry analysis by relaxing the weighting
constraints taken in \cite{sak,olv}. We comment on sufficiency of
having a Lax pair to be integrable in the symmetry sense.

\section{Construction of Recursion Operators}

From here on, by nc-Burgers equation we shall specifically refer
to the right-handed nc-Burgers equation
\begin{equation}
 u_{t}=u_{2x}+2uu_{x} \label{rncB}
\end{equation}
and present the explicit results pertaining to this version only. This is because all
 the results given here correspond to that of left-handed nc-Burgers equation by an
 interchange of left multiplication operator $L_{\psi}(\phi) =\psi \phi$ with the
 right multiplication $R_{\psi}(\phi) = \phi \psi$ at their every occurrence.

The Lax representation for the nc-Burgers hierarchy with
\begin{equation}
{\cal L} = D_{x}+L_{u}. \label{lx}
\end{equation}
is given by
\begin{equation}
  {\cal L}_{t_{n}}=[{\cal A}_{n}, {\cal L}], \label{lxh}
\end{equation}
where ${\cal A}_{n}=( {\cal L}^{n})_{\ge 1}$ and $D_{x}$ denotes the total derivative with respect to $x$.

When a hierarchy of Lax pairs is known, a technique for constructing  recursion operator
for the associated symmetry hierarchy is given in  \cite{m1}. This technique is based mainly
on the identification of the relation between the Lax representations of the individual symmetries
in the hierarchy. For the detailed explanation of the method and explicit examples  we refer to \cite{m1}.

The relation among the Lax representations of the individual
equations (symmetries) in the nc-Burgers hierarchy is
\begin{equation}
 {\cal L}_{t_{n+1}}={\cal L}{\cal L}_{t_{n}} + [{\cal T}_{n},{\cal L}].
\end{equation}
With the ansatz for the remainder ${\cal T}_{n}=a_{n}D_{x}+b_{n}$,
solution of this operator equality gives $a_{n}=ad_{{\cal L}}^{-1}
u_{n}$,and arbitrary  $b_{n}$. Therefore we have  the following
recursion formula
\begin{equation}
 u_{n+1}=(D_{x} + L_{u})u_{n}+R_{u_{x}}ad_{{\cal L}}^{-1}u_{n}-ad_{{\cal L}}b_{n} \label{recfor}
\end{equation}
among the successive symmetries. Here  $ad_{{\cal L}}=D_{x}+L_{u}-R_{u}$.

Considering the vector space structure of symmetries on which the
recursion operators act, we first choose $b_{n}=0$ and arrive at
the time independent recursion operator of the nc-Burgers equation
which is
\begin{equation}
 {\cal R}_{1}=D_{x}+L_{u}+R_{u_{x}}ad_{{\cal L}}^{-1}.
\end{equation}
To show that ${\cal R}_{1}$ is a conventional recursion operator
for the nc-Burgers hierarchy, one needs to prove that it satisfies
the basic defining condition \cite{olv1}
\begin{equation}
 {\cal R}_{t}=[{\cal F},{\cal R}]
\end{equation}
where ${\cal F}=D_{x}^{2}+2L_{u}D_{x}+2R_{u_{x}}$ is the Frechet
derivative of the right hand side of nc-Burgers
equation~(\ref{rncB}). In the present case however, it is more
convenient to verify this basic condition in its equivalent form
\cite{kar}
\begin{equation}
 {\cal M}_{t}- {\cal F}{\cal M}={\cal M}{\cal N}^{-1}({\cal N}_{t}-{\cal F}{\cal N}) \label{r1}
\end{equation}
where ${\cal R}_{1}={\cal M}{\cal N}^{-1}$ with
\begin{equation}
 {\cal M}=(D_{x}+L_{u})ad_{{\cal L}}+R_{u_{x}},~~~~~{\cal N}=ad_{{\cal L}}.
\end{equation}
\noindent It can be straightforwardly verified that Eqn.(\ref{r1})
indeed holds and therefore ${\cal R}_{1}$ is a recursion operator
for nc-Burgers equation. It generates an infinite hierarchy of
symmetries if an initial one is given. The first few symmetries
starting from $\sigma_{0}=u_{x}$ are
\begin{eqnarray}
\sigma_{1}&=&u_{2x}+2uu_{x}, \nonumber \\
\sigma_{2}&=&u_{3x}+3uu_{2x}+3{u_{x}}^{2} +3u^{2}u_{x}, \nonumber\\
\sigma_{3}&=&u_{4x}+4uu_{3x}+4u_{x}uu_{x}+4u_{2x}u_{x}+6u^{2}u_{2x}
+4u^{3}u_{x}+6u_{x}u_{2x} \nonumber \\
&&+8u{u_{x}}^{2}. \nonumber
\end{eqnarray}
As in the commutative Burgers equation, there is another recursion
operator of the nc-Burgers equation which is explicitly time
dependent. We can determine the time dependent recursion operator
${\cal R}_{2}$, with the help of the $b_{n}$, by inspection. Hence
choosing $b_{n}=(tD_{x}+tL_{u}+\frac{x}{2})ad_{{\cal
L}}^{-1}u_{n}$ in (\ref{recfor}) we get the time dependent
recursion operator
\begin{equation}
 {\cal R}_{2}=ad_{{\cal L}} (tD_{x}+tL_{u}+\frac{x}{2})ad_{{\cal L}}^{-1}.
\end{equation}
\noindent Again, as in the commutative case, this time-dependent
recursion operator ${\cal R}_{2}$ is a weak recursion operator
meaning that despite satisfying the basic condition~(\ref{r1}),
this recursion operator fails to generate higher symmetries
correctly. Going through the algorithm \cite{san,m2}, the
corrected time-dependent recursion operator for the nc-Burgers
equation is obtained to be
\begin{equation}
 {\cal R}_{2}=ad_{{\cal L}} (tD_{x}+tL_{u}+\frac{x}{2})ad_{{\cal L}}^{-1}+\frac{1}{2}(\frac{1}{2}\
+tL_{u_{x}})D^{-1}_{t}\Pi ad_{{\cal L}},
\end{equation}
where $\Pi$ is the projection operator defined as $\Pi h(t,x,u,u_{x},...)=h(t,0,0,...)$ for any function $h$ .
\noindent
The first few symmetries are
\begin{eqnarray}
\sigma_{0}&=&\frac{1}{2}+t u_{x}, \nonumber \\
\sigma_{1}&=&t^{2}(u_{2x}+2uu_{x})+t(u+xu_{x})+\frac{1}{2}x
, \nonumber \\
\sigma_{2}&=&t^{3}(u_{3x}+3uu_{2x}+3{u_{x}}^{2} +3u^{2}u_{x})
\nonumber\\
&+&t^{2}(3u_{x}+\frac{3}{2} xu_{2x}+3xuu_{x}+\frac{3}{2} u^{2})
\nonumber \\
&+&t(\frac{3}{4}+\frac{3}{4} x^{2} u_{x}+\frac{3}{2}ux)+\frac{3}{8} x^{2}.
\end{eqnarray}
\vspace{0.3cm}

\section{Mixed nc-Burgers equations}
The mixed nc-Burgers equation having an arbitrary constant
$\alpha$
\begin{equation}
v_{t}=v_{2x}+(\alpha-2)v_{x}v + \alpha vv_{x}\label{mncB}
\end{equation}
was introduced with the Lax pair
\begin{equation}
{\cal L}_{mixed}=D_{x}+v,\;\; {\cal A}_{mixed}=D^{2}_{x}+2vD_{x} + 2v_{x} + \alpha v^{2}
\label{lxm}
\end{equation}
in \cite{ham}. Even though the mixed nc-Burgers
equation~(\ref{mncB}) admits the particular Lax
formulation~(\ref{lxm}) which may be regarded as indicating
integrability. This equation has not shown up in symmetry
classifications \cite{sak,olv}. Moreover, we have already obtained
the recursion operators that  ${\cal L}_{mixed}$ can give. They
are only the recursion operators of the nc-Burgers equation
(\ref{rncB}) admitting Lax representations ${\cal
L}_{mixed}=D_{x}+L_{u}$ or ${\cal L}_{mixed}=D_{x}+R_{u}$.

Therefore, here we reinvestigate a slightly generalized mixed
nc-Burgers equations for higher symmetry. In our specific attempt,
we  relax the weighting scheme used in \cite{sak,olv} for the
selection of the terms to be included in the candidate symmetry.
We included all the terms with polynomial and derivative orders up
to four. We have the following proposition.

{\bf Proposition:} {\em The equation of form
\begin{equation}
u_{t}=u_{2x} + auu_{x} + bu_{x}u \label{gmix}
\end{equation}
with $a,b \in {\mathbb {R}}$ and $ab \neq 0$, $u$ is is
non-commutative,

i) does not admit any higher symmetry from the class of equations
\begin{eqnarray}
 u_{t}&=&\nu(t,x)+\sum_{i=0}^{4}\alpha^{i}(t,x)u_{ix}+\sum_{i,j=0}^{4}\beta^{ij}(t,x)u_{ix}u_{jx}
+\sum_{i,j,k=0}^{4}\gamma^{ijk}(t,x)u_{ix}u_{jx}u_{kx}\nonumber\\
&+&\sum_{i,j,k,l=0}^{4}\delta^{ijkl}(t,x)u_{ix}u_{jx}u_{kx}u_{lx},\nonumber
\end{eqnarray}

ii) admits only the Lie-point symmetries
\begin{eqnarray}
\sigma_{1}&=&u_{t}, \nonumber \\
\sigma_{2}&=&u_{x}, \nonumber\\
\sigma_{3}&=&2tu_{t}+xu_{x}+u,\nonumber \\
\sigma_{4}&=&1+(a+b)tu_{x}, \nonumber
\end{eqnarray}

iii) in particular, when $b=-a$, $\sigma_{1}$, $\sigma_{2}$, $\sigma_{3}$ remains as they are but $\sigma_{4}=1$
generalizes to $\sigma_{4}=h(t,x)$ where $h(t,x)$ is a solution of $h_{t}=h_{xx}$.}

So, to the extent of the above proposition, integrable versions of
nc-Burgers equations are only the left- and right-handed ones
which are already given in \cite{sak}. We claim that there does
not exist either a Lax hierarchy or a recursion operator for the
mixed nc-Burgers equation~(\ref{mncB}).

Nevertheless, a non-integrable equation with a Lax formulation is
not unusual. Such (commutative) equations are shown to exist in
\cite{cal-nucci} and further investigated in \cite{sergei}.

One of the authors A.K. is thankful to Sergei Sakovich for various valuable discussions.
This work is partially supported by the Scientific and Technological
Research Council of Turkey (TUBITAK) and  Turkish Academy of Sciences (TUBA).

\end{document}